\documentstyle[twocolumn,epsfig,graphicx]{article}
\newcommand{\beq}{\begin{equation}}
\newcommand{\eeq}{\end{equation}}
\newcommand{\bea}{\begin{eqnarray}}
\newcommand{\eea}{\end{eqnarray}}

\def\nuc#1#2{\relax\ifmmode{}^{#1}{\protect\text{#2}}\else${}^{#1}$#2\fi}

\title{\large 
PROTON-INDUCED FISSION CROSS SECTION CALCULATION WITH 
THE LANL CODES CEM2K+GEM2 AND LAQGSM+GEM2
} 
\vspace{1.2cm}
\author{
Mircea I. Baznat, Konstantin K. Gudima
\\
{\it 
Institute of Applied Physics, Academy of Science of Moldova,
Chisinau, MD-2023, Moldova}\\
\\
Stepan G. Mashnik\\
{\it Los Alamos National Laboratory,  Los Alamos, NM 87545, USA}\\
}
\begin{document}
\maketitle

\noindent ABSTRACT
\vspace{0.1cm}

\noindent

The improved Cascade-Exciton Model code CEM2k and the Los Alamos
version of the Quark-Gluon String Model code LAQGSM, previously merged 
with the Generalized Evaporation Model code of Furihata (GEM2) were
further modified to provide reliable proton-induced fission
cross sections for applications. By adjusting two parameters in 
GEM2 for each measured reaction, we were able to describe very well
with CEM2k+GEM2 and LAQGSM+GEM2
 all available experimental fission cross sections
induced by protons with energies from 20 MeV to 10 GeV both for
subactinide and actinide targets.
We also successfully tested our approach on several reactions
induced by neutrons, pions, and photons.

\section*{Introduction}
\noindent

In recent years, an improved version of the Cascade-Exciton Model (CEM), 
contained in the code CEM2k \cite{CEM2k}
and the Los Alamos version of the Quark-Gluon String Model,
implemented in the high-energy code LAQGSM \cite{LAQGSM}
have been developed at the Los Alamos National Laboratory
for a number of applications.
CEM2k is intended to describe nucleon-,
pion-, and photon-induced reactions at incident energies up to about 5 GeV,
while LAQGSM  describes both particle- and nucleus-nucleus reactions
at energies up to about 1 TeV/nucleon.
Originally, both CEM2k and LAQGSM were not able to describe fission reactions 
and production of light fragments heavier than $^4$He, as they had neither 
a high-energy-fission nor a fragmentation model. 
Recently, we addressed these problems \cite{SATIF6,SantaFe02}
by further improving
our codes and by merging them with the Generalized Evaporation Model
code GEM2 developed by Furihata \cite{GEM2}. 

GEM2 is an extension by Furihata 
of the Dostrovsky {\it et al.} \cite{Dostrovsky} evaporation model
as implemented in LAHET \cite{LAHET}
to include up to 66 types of particles and fragments that
can be evaporated from an excited compound nucleus plus a modification
of the version of Atchison's fission model \cite{RAL}
used in LAHET. It was found \cite{SATIF6,SantaFe02} that if we were to
merge GEM2 with CEM2k or LAQGSM without any modifications, the new code
would not describe correctly the fission cross section (and the yields of
fission fragments). This is because Atchison
fitted the parameters of his fission model when it was coupled 
with the Bertini Intra-Nuclear Cascade (INC) 
\cite{BertiniINC} which differs
from our INC. In addition, Atchison did not model 
preequilibrium emission. Therefore, the distributions of fissioning
nuclei in $A$, $Z$, and excitation energy $E^*$ simulated by Atchison
differ significantly of the distributions we get; 
as a consequence, all the fission characteristics are also different.
Furihata used GEM2 coupled either with the Bertini INC 
\cite{BertiniINC} or with
the ISABEL \cite{ISABEL} INC code, which also differs from our INC, and did 
not include preequilibrium particle emission. Therefore the
real fissioning nuclei simulated by Furihata differ from the ones in
our simulations, and the parameters adjusted by Furihata to work the best 
with her INC will not be the best for us. To get a good description 
of fission cross sections (and fission-fragment yields)
we need to modify at least two parameters in GEM2 (see details in
\cite{SATIF6,SantaFe02}). This
problem was solved both for CEM2k+GEM2 and LAQGSM+GEM2 
in the present work.

\section*{Calculation of $\sigma_f$  in GEM2}

A comprehensive description of GEM2 was published by Furihata
\cite{GEM2}, some details may be found in our papers 
\cite{SATIF6,SantaFe02},
therefore we recall here only how 
fission cross sections are calculated by GEM2, as we need to
modify them here.
The fission model used in GEM2 is based on Atchison's model \cite{RAL},
often referred in the literature as the Rutherford Appleton Laboratory (RAL)
model, which is where Atchison developed it. There are two choices of 
parameters for the fission model: one of them is the original parameter 
set by Atchison \cite{RAL} as implemented in LAHET \cite{LAHET}, 
and the other is a parameter set evaluated by Furihata \cite{GEM2},
used here as a default of GEM2.

The Atchison fission model is designed to only describe fission of
nuclei with $Z \geq 70$
(we extended it in our codes down to $Z \geq 65$). 
It assumes that fission competes only with
neutron emission, {\it i.e.}, from the widths $\Gamma_j$ of n, p, d, 
t, $^3$He, and $^4$He,
the RAL code calculates the probability of evaporation of any 
particle. When a charged particle is selected to be evaporated, 
no fission competition is taken into account. When a neutron is
selected to be evaporated, the code does not actually simulate its 
evaporation,
instead it considers that fission may compete,
and chooses either fission or evaporation of a neutron according to
the fission probability $P_f$. This quantity is treated by the RAL code 
differently for the elements above and below $Z=89$. 

1) $70 \leq Z_j \leq 88$. For fissioning nuclei with $70 \leq Z_j \leq 88$,
GEM2 uses the original Atchison calculation of the neutron emission
width $\Gamma_n$ and fission width $\Gamma_f$ to estimate the fission
probability as
\beq
P_f = {\Gamma_f \over {\Gamma_f+\Gamma_n}}={1 \over {1+\Gamma_n/\Gamma_f}}.
\eeq
Atchison uses \cite{RAL} the Weisskopf and Ewing statistical model
\cite{Weisskopf}
with an energy-independent pre-exponential factor for the level density 
and Dostrovsky's \cite{Dostrovsky} inverse cross section for neutrons
and estimates the neutron width $\Gamma_n$ as
\bea
\Gamma_n = 0.352 \bigl(1.68 J_0 + 1.93A_i^{1/3}J_1 \nonumber \\
+ A_i^{2/3}(0.76J_1 - 0.05 J_0)\bigr),\nonumber 
\eea
where $J_0$ and $J_1$ are functions of the level density parameter $a_n$
and $s_n (=2\sqrt{a_n(E-Q_n-\delta)})$ as
$$J_0 = { {(s_n-1) e^{s_n} + 1} \over {2 a_n} },$$
$$J_1 = { {(2s_n^2 - 6s_n + 6) e^{s_n} + s_n^2 -6} \over {8a_n^2} }.$$
The RAL model uses
a fixed value for the level density parameter $a_n$, namely
$$
a_n = (A_i - 1) /8 .
$$
The fission width for nuclei with $70 \leq Z_j \leq 88$ is calculated 
in the RAL model and in GEM2 as
$$
\Gamma_f = { {(s_f - 1)e^{s_f} + 1} \over a_f },
$$
where $s_f = 2 \sqrt{a_f (E-B_f - \delta)}$ and the level density parameter
in the fission mode $a_f$ is fitted by Atchison to describe the measured
$\Gamma_f / \Gamma_n$ as
\beq
a_f = a_n \Bigl(1.08926 + 0.01098 ( \chi - 31.08551)^2\Bigr),
\eeq
and $\chi = Z^2/A$.

2) $Z_j \geq 89$. For heavy fissioning nuclei with $Z_j \geq 89$
GEM2 follows the RAL model and does not calculate at all
the fission width $\Gamma_f$ and does not use Eq.\ (1) to estimate
the fission probability $P_f$. Instead, the following semi-empirical
expression obtained by Atchison by approximating the experimental
 values of
$\Gamma_n / \Gamma_f$ published by
Vandenbosch and Huizenga \cite{Vandenbosch}  is used to calculate
the fission probability:
\beq
\log (\Gamma_n / \Gamma_f ) = C(Z_i) ( A_i - A_0(Z_i)),
\eeq
where $C(Z)$ and $A_0(Z)$ are constants dependent on the nuclear charge $Z$
only. The values of these constants are those
used in the current version of LAHET \cite{LAHET} and are tabulated in Table 1
(note that some adjustments of these values have been done since
Atchison's papers \cite{RAL} were published).  

\begin{center}
Table 1. $C(Z)$ and $A_0(Z)$ values used in GEM2

\vspace{2mm}
\begin{tabular}{|c|c|c|}
\hline \hline
\hspace{5mm} Z\hspace{5mm} &\hspace{5mm} $C(Z)$\hspace{5mm} &\hspace{5mm}
 $ A_0(Z)$\hspace{5mm} \\
\hline
89 & 0.23000 & 219.40\\
90 & 0.23300 & 226.90\\
91 & 0.12225 & 229.75\\
92 & 0.14727 & 234.04\\
93 & 0.13559 & 238.88\\
94 & 0.15735 & 241.34\\
95 & 0.16597 & 243.04\\
96 & 0.17589 & 245.52\\
97 & 0.18018 & 246.84\\
98 & 0.19568 & 250.18\\
99 & 0.16313 & 254.00\\
100& 0.17123 & 257.80\\
101& 0.17123 & 261.30\\
102& 0.17123 & 264.80\\
103& 0.17123 & 268.30\\
104& 0.17123 & 271.80\\
105& 0.17123 & 275.30\\
106& 0.17123 & 278.80\\
\hline \hline
\end{tabular}
\end{center}

\section*{Prokofiev's Approximation of $\sigma_f$}

We choose not to use in the present work experimental fission
cross sections directly as they are published in the literature. 
Fig. 1 (kindly provided by Dr. Prokofiev) explains well the reason:
The point is that for intermediate- and high-energy reactions,
where our codes are supposed to be used, the experimental data on
proton-induced fission cross sections are sparse and not as
precise as for low-energy reactions measured for reactor applications.
Intermediate- and high-energy experimental fission cross sections
induced by neutrons, pions, and other projectiles are even more sparse
than the ones measured with protons. As one can see from Fig. 1,
fission cross sections  measured at such energies
in different experiments differ so significantly from each other that 
it is difficult to use such
data in development and validation of models and codes, without
a special analysis of all details of every measurement. 
Fortunately, this has been done
by Prokofiev \cite{Prokofiev} so we use here his results.
Prokofiev spent many years on compiling proton-induced measured 
fission cross sections and on analyzing the details of each experiment.
As a result, he divided all measurements into three categories:
1) the highest, where obtained data are very reliable and can be used 
without any mistrust;
2) high-quality data, reliable, but requiring some normalization;
3) data of low reliability, that would be better not used.
Then, using only measurements from the first group and data from
the second group after a corresponding re-normalization, Prokofiev
developed systematics for proton-induced fission cross sections
for all preactinide and actinide nuclei for which he was able to find 
enough data \cite{Prokofiev,Prokofiev2}. 
At our energies, we consider Prokofiev's systematics as the most reliable
``experimental" fission cross sections and prefer to use them
to develop and test our codes instead of using experimental values
published in original publications by different authors.

For subactinide nuclei from $^{165}$Ho to $^{209}$Bi and incident proton
energies above 70 MeV,
Prokofiev proposed \cite{Prokofiev} the following universal parameterization
for the proton-induced fission cross section, $\sigma_f(E_p)$ [mb]:
\bea
\sigma_f(E_p) &=& P_1 \{1 - \exp [ - P_3 (E_p - P_2) ] \} \nonumber \\
&\times& (1 - P_4 \ln E_p) ,
\eea
where $E_p$ is the incident proton energy [MeV] and $P_1$, $P_2$, $P_3$,
and $P_4$ are fitting parameters.
$P_4$ was fitted as
\beq
P_4(Z^2/A) = \cases{ 0 &if  $Z^2/A \leq 32.32$,\cr
   Q_{4,1} + Q_{4,2} Z^2/A&if $Z^2/A > 32.32$,\cr}
\eeq
where fitting parameters $Q_{ij}$ are given in Table 2.
Parameters $P_1$, $P_2$, and $P_3$ were fitted as
\beq
P_i(Z^2/A) =
\exp [ Q_{i,1} + Q_{i,2} (Z^2/A) + Q_{i,3} (Z^2/A)^2 ] .
\eeq

\vspace*{2mm}
\begin{center}
Table 2. Parameters $Q_{ij}$ in the $P_i(Z^2/A)$ systematics
for target nuclei from Ho to Bi \cite{Prokofiev}

\vspace{2mm}
\begin{tabular}{|c|c|c|c|}
\hline 
$i$ & $j=1$ & $j=2$ & $j=3$ \\
\hline
1 & 119.0  & -7.852  & 0.1332\\
2 & 9.979  & -0.1847 & 0     \\
3 & -27.40 & 0.6792  & 0     \\
4 & -1.140 & 0.0352  & 0 \\
\hline 
\end{tabular}
\end{center}

For actinide nuclides from $^{232}$Th to $^{239}$Pu and incident proton
energies above 20 MeV, 
Prokofiev found \cite{Prokofiev} $P_2 = 12.1$, $P_3 = 0.111$, $P_4 = 0.067$,
and
\beq
P_1(Z^2/A) = R_{11} \{1 - \exp [ - R_{13}(Z^2/A - R_{12} ) ] \} ,
\eeq
where $R_{11} = 2572$, $R_{12} = 34.99$, and $R_{13} = 2.069$.
Numerical values of all $P_i$ parameters of the nuclear targets fitted
by Prokofiev together with the energy interval of fitting are published
in Tab. 4 of Ref. \cite{Prokofiev}.

In Ref. \cite{Prokofiev2}, Prokofiev extended his systematics
to describe fission cross sections of preactinide nuclei from
$^{197}$Au to $^{209}$Bi in the energy region from 35 to 70 MeV 
and to predict fission cross sections for nuclei between $^{209}$Bi
and $^{232}$Th, where not a single data point is available at present.
It was found \cite{Prokofiev2} that one can approximate fission cross
sections of preactinides between Au and Bi at proton energies
between 35 and 70 MeV with the formula
\beq
\sigma(E_p) = \sigma_0 \exp \Bigl[ - { {(E_p - E_0)^2} 
\over {2 w^2} }\Bigr] ,
\eeq
where $E_0 = 76.3$ MeV.
Parameters $w$ and $\sigma_0$ depend
on the fissioning system and characterize, respectively,
the steepness and the absolute scale of the fission excitation functions
and are approximated as following:
\beq
w(A,Z) = a + b(Z^2/A) + c\; \delta W_{gs}(A,Z) ,
\eeq
where $\delta W_{gs}$ is the shell correction to the ground-state
mass of the fissioning nucleus calculated using the systematics
of Myers and Swiatecki \cite{MyersSwiatecki}, and
$a = -33.667$, $b = 1.5699$, and $c = 0.30069$. 
Parameter $\sigma_0$ was fitted as
\beq
\sigma_0 = \sigma_b \exp \Bigl[ { {(E_b - E_0)^2} 
\over {2 w^2} }\Bigr] ,
\eeq
where $E_b = 70$ MeV and $\sigma_b = \sigma(E_b)$ is
calculated according to the high-energy systematics given by Eq. (4).

To predict fission cross sections for nuclei between $^{209}$Bi and
$^{232}$Th at proton energies above 70 MeV
were there are no data, it was suggested \cite{Prokofiev2} 
that parameters $P_i$ of Eq. (4) can be found by interpolation 
of the systematics \cite{Prokofiev} predictions. The logarithmic interpolation
scheme was chosen \cite{Prokofiev2}:
\beq
\ln P_i = C_{i1} + C_{i2} \: x ,
\eeq

\begin{figure*}[!ht]\begin{center}
\vspace{-2.5cm}

\centerline{\epsfxsize 13.8cm \epsffile{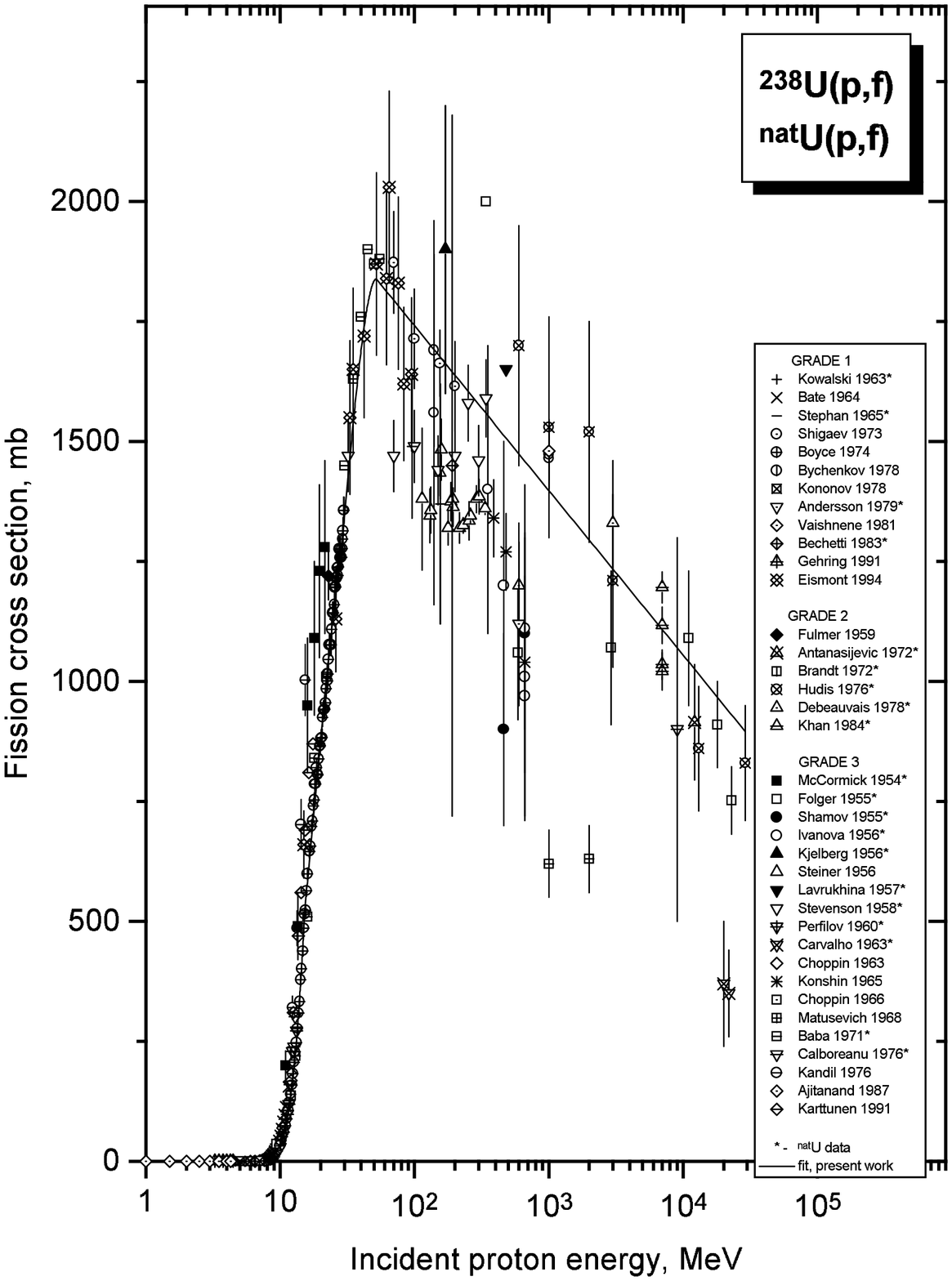}} 

\vspace{-1.8cm}

\caption{Experimental proton-induced fission cross sections
of $^{238}$U and $^{nat}$U nuclei compiled by Prokofiev (symbols)
compared with results of his sytematics \cite{Prokofiev}
for these cross sections (line). We thank Dr. Prokofiev for
sending us this figure.
}\label{fig1} 
\end{center}
\end{figure*}

{\noindent
where the constants $C_{ij}$ ($i = 1 \dots 4, j = 1, 2$) are calculated
as following:}
\beq
C_{i1} = { {x_{Th} \ln P_i (x_{Bi}) - x _{Bi} \ln P_i (x_{Th}) } 
\over {x_{Th} - x_{Bi} } } ,
\eeq
\beq
C_{i2} = { {\ln P_i (x_{Th}) - \ln P_i (x_{Bi}) }
\over {x_{Th} - x_{Bi} } } ,
\eeq
where $P_i(x)$ are predictions of the systematics \cite{Prokofiev}
described by Eqs. (5-7), and indexes ``Bi" and ``Th" denote the
$^{209}$Bi+p and $^{232}$Th+p fissioning systems, correspondingly.
The resulting $C_{ij}$ values are: 
$C_{11} = -27.74$, $C_{12} =0.9906$, $C_{21} = 25.83$, $C_{22} =-0.6567$,
$C_{31} = -45.80$, $C_{32} = 1.227 $, $C_{41} = -10.95$,
and $C_{42} = 0.2320$. Bellow, we use values provided by Eqs. (4-13)
to adjust the calculation of fission cross sections in
our CEM2k+GEM2 and LAQGSM+GEM2 codes.

\section*{Results}

The main parameters that determine the fission cross sections
calculated by GEM2 are the level density parameter in the
fission channel, $a_f$ (or more exactly, the ratio $a_f/a_n$
as calculated by Eq. (2)) for preactinides, and parameter
$C(Z)$ in Eq. (3) for actinides. The sensitivity of results to
these parameters is much higher than to fission
barriers used in calculation or other parameters of the model.
Therefore we choose to adjust only these two parameters in our
merged CEM2k+GEM2 and LAQGSM+GEM2 codes. We do not change
the form of systematics (2) and (3) derived by Atchison.
We only introduce here additional coefficients both to $a_f$ and $C(Z)$,
replacing $a_f \to C_a \times a_f$ in Eq. (2) and 
$C(Z_i) \to C_c \times C(Z_i)$
in Eq. (3) and fit $C_a$ and $C_c$ both for CEM2k+GEM2 and
LAQGSM+GEM2 codes for all nuclei and incident proton energies where
Prokofiev's systematics apply. No other parameters in GEM2 or
our CEM2k and LAQGSM were changed. 
For preactinides, we had to fit only $C_a$.
The values of $C_a$
found by fitting our results to Prokofiev's predictions 
are close to one and change smoothly with changing the
proton energy and the charge or mass number of the target.
Such finding gives us confidence in our procedure, and
allows us to interpolate or extrapolate the values of $C_a$
for nuclei and incident proton energies not covered by
Prokofiev's systematics.
For actinides, as described in \cite{SATIF6,SantaFe02},
we have to fit both $C_a$
and $C_c$. The values of $C_a$ we find are also very close to one,
while the values of $C_c$ are more varied, but both of them
change smoothly with the proton energy and Z or A of the   
target, that again allows us to interpolate and extrapolate
them for nuclei and energies outside Prokofiev's systematics.

We fixed the fitted values of $C_a$ and $C_c$ in data blocks 
in our codes and complemented them with routines for their
interpolation/extrapolation outside the region covered by
Prokofiev's systematics. We believe that such a procedure 
provides quite a reliable fission cross section calculation
by our codes, at least for proton energies and target-nuclei
not too far from the ones covered by Prokofiev's systematics.
Our results by CEM2k+GEM2 for preactinides are shown in Fig. 2,
and for actinides, in Fig. 3. Results by LAQGSM+GEM2 are very similar,
almost coinciding with the ones shown in Figs. 2 and 3, therefore we
do not duplicate them here. One can see that after fitting $C_a$
and $C_c$, the fission cross sections calculated by our codes
reproduce very well all the experimental data covered by Prokofiev's
systematics.

To see how this approach works for reactions induced by other
projectiles, we tested our codes on several reactions induced by
neutrons, pions, and photons, without any more changes or fitting.
Fig. 4 shows several examples of such results.
We see that our codes describe them from quite well to very well,
although experimental data on pion-induced
fission cross sections are not so rich and precise,
and it is difficult to draw 
conclusions from a comparison to this data.
The fact that we give such fits to fission induced by other probes 
gives us confidence in the value of
the fitting procedure we performed in our CEM2k+GEM2
and LAQGSM+GEM2 codes.

\section*{Acknowledgment}
\noindent
We thank Dr.\ Prokofiev for useful discussions,
collaboration, and for sending us his figures
with experimental  and approximated
proton-induced fission cross sections and
Drs.\ Sierk and Prael for numerous discussions
and help.
We are grateful to Prof.\ Peterson for sending
us his compilation of experimental pion-induced fission cross sections.
The work has been supported by the U.S. Department of Energy 
and by the Moldovan-U.\ S.\ Bilateral Grants Program, CRDF Project MP2-3025.
S.G.M.\ acknowledges partial support from a NASA
Astrophysics Theory Program grant.

\newpage
\begin{figure*}[!ht]\begin{center}
\vspace{-6.0cm}
\hspace*{-3cm}
\includegraphics[width=21.5cm]{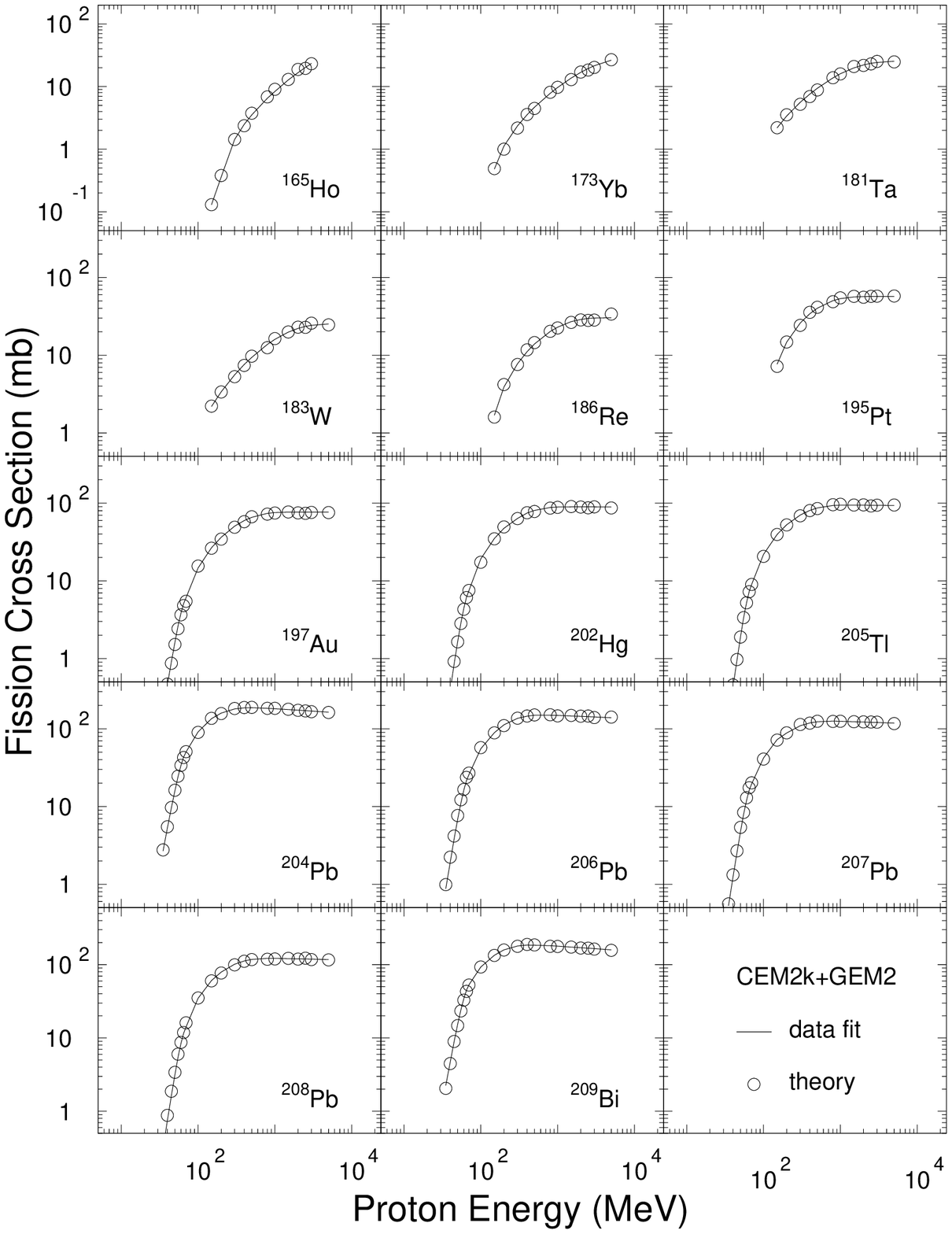}

\vspace{-3cm}
\caption{Comparison of Prokofiev's \cite{Prokofiev,Prokofiev2}
sytematics of experimental (p,f) cross sections of
$^{165}$Ho, $^{173}$Yb, $^{181}$Ta, $^{183}$W, $^{186}$Re, $^{195}$Pt, 
$^{197}$Au, $^{202}$Hg, $^{205}$Tl, $^{204}$Pb, $^{206}$Pb, $^{207}$Pb, 
$^{208}$Pb, and $^{209}$Bi nuclei (lines)
with our present CEM2k+GEM2 calculations
(circles).
}\label{fig2}
\end{center}
\end{figure*}

\newpage
\begin{figure*}[!ht]\begin{center}

\vspace{-8.0cm}
\hspace*{-3cm}
\includegraphics[width=21.5cm]{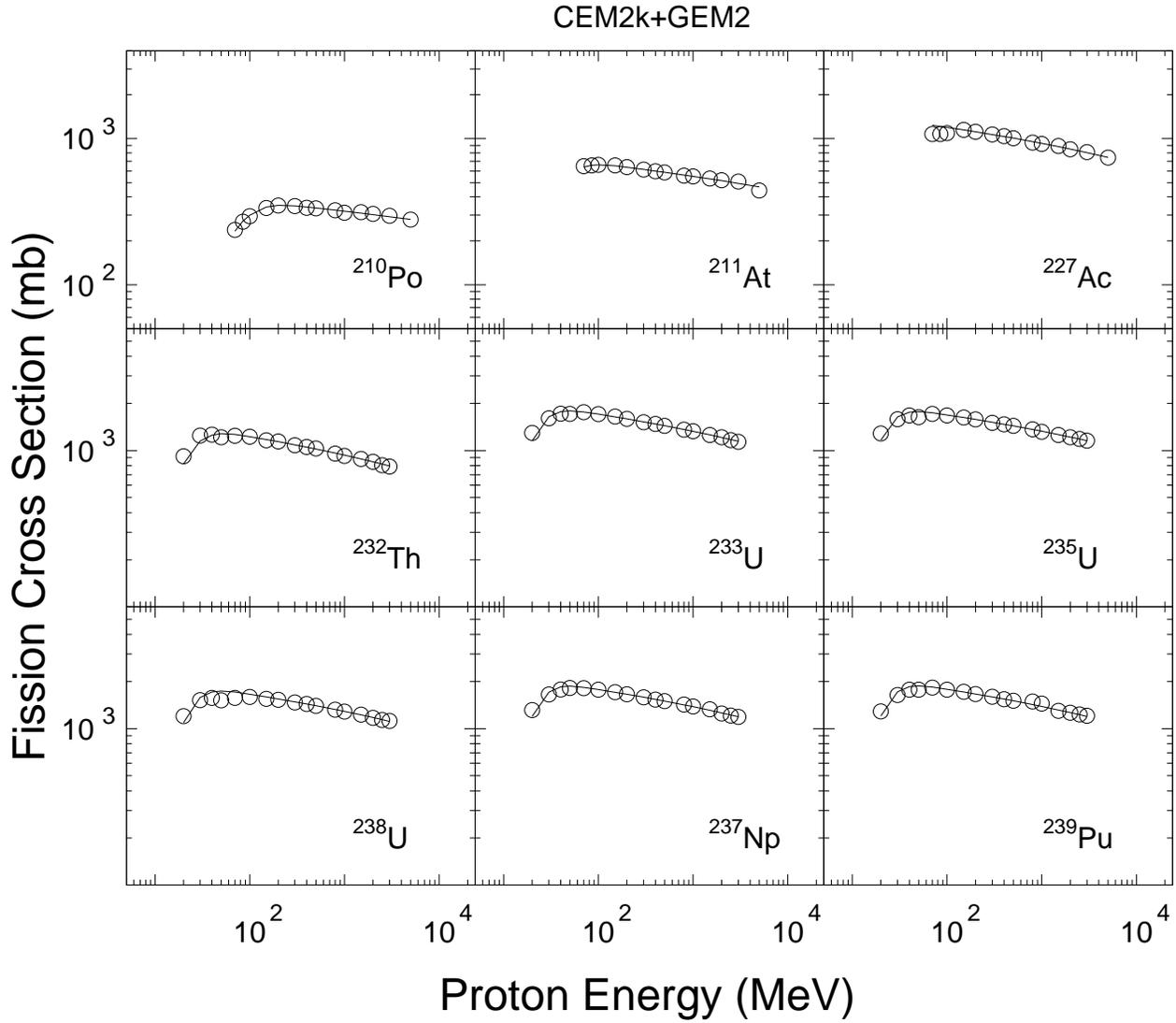}

\vspace{-10.5cm}
\caption{
Comparison of Prokofiev's \cite{Prokofiev}
sytematics of experimental (p,f) cross sections of
$^{232}$Th, $^{233}$U, $^{235}$U $^{238}$U, $^{237}$Np, and $^{239}$Np nuclei
and of predicted \cite{Prokofiev2} (p,f) cross sections for
$^{210}$Po, $^{211}$At, and $^{227}$Ac targets (lines)
with our present CEM2k+GEM2 calculations
(circles).
}\label{fig3}
\end{center}
\end{figure*}

\newpage
\begin{figure*}[!ht]\begin{center}

\vspace{-3.0cm}
\hspace*{-0.5cm}
\includegraphics[width=19.5cm]{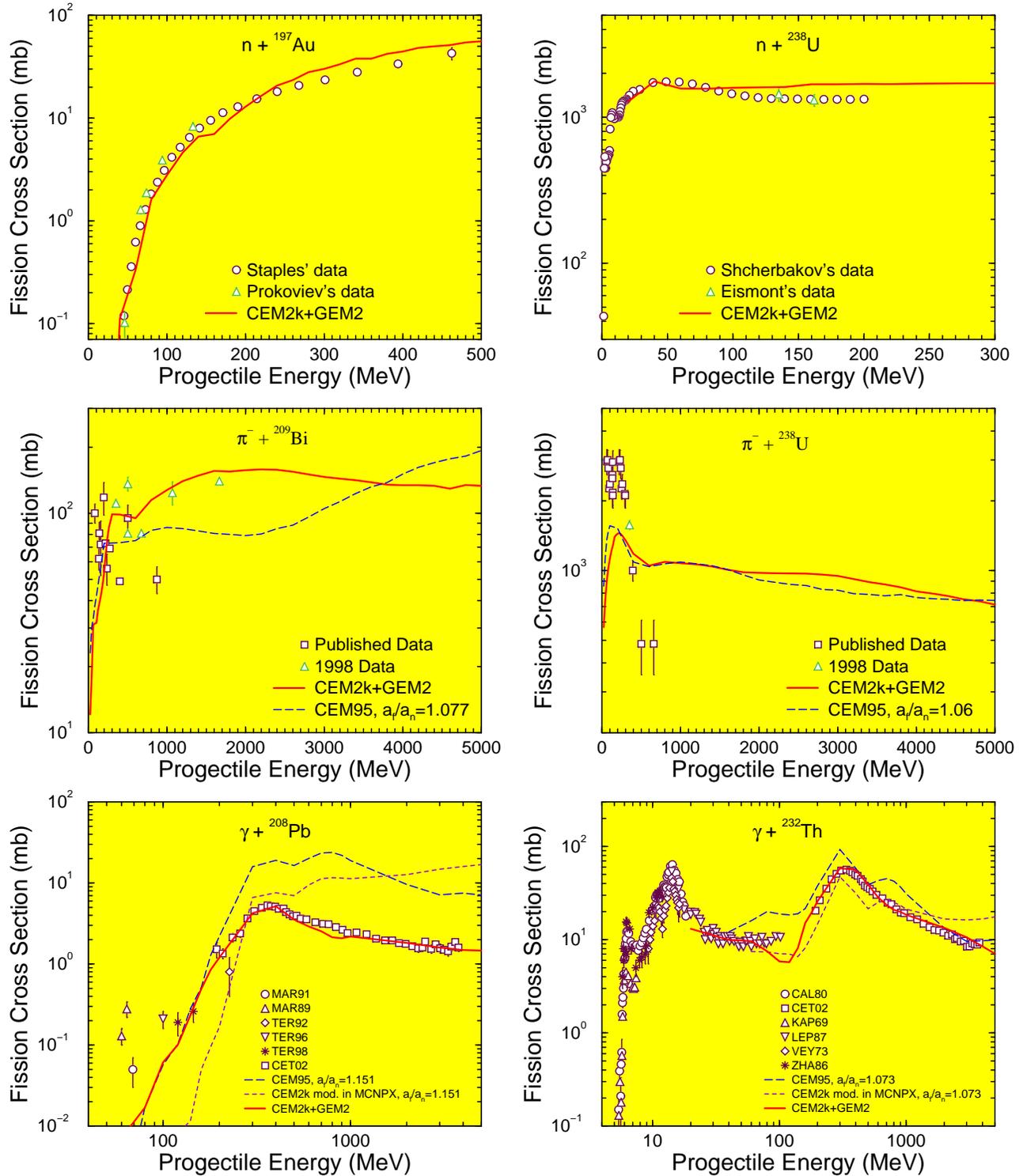}

\vspace{-4.3cm}
\caption{
Comparison of calculated by the modified here CEM2k+GEM2
code fission cross sections
induced by neutrons on $^{197}$Au and $^{238}$U, 
$\pi^-$ on $^{209}$Bi and $^{238}$U, and 
$\gamma$ on $^{208}$Pb and $^{232}$Th with experimental
data and results by previous versions of CEM (see details and
references in
\cite{CEM2k}), as indicated. Experimental data are from: 
1) n: Staples \cite{Staples}, Prokofiev \cite{Prokofievdata};
Shcherbakov \cite{Shcherbakov}, Eismont \cite{Eismont};
2) $\pi^-$: \cite{Peterson};
3) $\gamma$: 
MAR91 \cite{MAR91},
MAR89 \cite{MAR89},
TER92 \cite{TER92},
TER96 \cite{TER96},
TER98 \cite{TER98},
CET02 \cite{CET02},
CAL80 \cite{CAL80},
KAP69 \cite{KAP69},
LEP87 \cite{LEP87},
VEY73 \cite{VEY73},
ZHA86 \cite{ZHA86}.
}
\end{center}
\end{figure*}

\end{document}